\documentclass[prb,twocolumn,superscriptaddress,floatfix,nopacs]{revtex4-1}
\usepackage{amsfonts,amsmath,epsf,dcolumn,natbib}
\usepackage{color}
\usepackage{graphicx}
\usepackage{multirow}
\usepackage{rotating}
\newcolumntype{.}{D{.}{.}{-1}}

\newcommand{\half}{\tfrac{1}{2}}

\newcommand{\be}{\begin{equation}}
\newcommand{\ee}{\end{equation}}
\newcommand{\bea}{\begin{eqnarray}}
\newcommand{\eea}{\end{eqnarray}}
\newcommand{\balg}{\begin{align}}
\newcommand{\ealg}{\end{align}}

\definecolor{purple}{rgb}{0.5, 0.0, 0.5}

\begin{document}
	\title{Deorbitalized meta-GGA Exchange-Correlation Functionals in Solids}
	\author{Daniel Mejia-Rodriguez}
	\email{dmejiarodriguez@ufl.edu}
	\affiliation{Quantum Theory Project, Department of Physics, University of Florida, Gainesville, FL 32611}
	\author{S.B.~Trickey}
	\email{trickey@qtp.ufl.edu}
	\affiliation{Quantum Theory Project, Departments of Physics and
		of Chemistry, University of Florida, Gainesville, FL 32611}
	
	\date{20 July 2018}

	\begin{abstract}
A procedure for removing explicit orbital dependence from
meta-generalized-gradient approximation (mGGA) exchange-correlation
functionals by converting them into Laplacian-dependent functionals
recently was developed by us and shown to be successful in molecules.
It uses an approximate kinetic energy density functional (KEDF)
parametrized to Kohn-Sham results (not experimental data) on a small
training set.  Here we present extensive validation calculations on
periodic solids that demonstrate that the same deorbitalization with
the same parametrization also is successful for those extended
systems.  Because of the number of stringent constraints used in its
construction and its recent prominence, our focus is on the SCAN
meta-GGA.  Coded in \textsc{vasp}, the deorbitalized version, SCAN-L, can be
as much as a factor of three  
faster than original SCAN, a potentially significant gain for large-scale
ab initio molecular dynamics.  
\end{abstract}

	\maketitle

\section{\label{Intro}Background and Motivation}

Accuracy, generality, and computational cost are competing priorities in 
the unrelenting search for theoretical constructs on which to base 
predictive condensed matter calculations. A critical problem is 
to predict the stable zero-temperature lattice structure of a crystal, 
its cohesive energy, its bulk modulus, and fundamental gap. Treatment of 
other physical properties 
(e.g., phonon spectra, transport coefficients, response functions, etc.) 
is, in principle at least, built upon ingredients drawn from solution
of that central problem.  

Beginning about four decades ago, the dominant paradigm which  
emerged for treating that central problem is density functional
theory (DFT) \cite{HK,Levy79,Lieb83} in its Kohn-Sham (KS) 
\cite{KS,EngelDreizlerBuch} form. For an $N_e$ electron system, the 
KS procedure recasts the DFT variational problem as one for a counterpart 
non-interacting system which has its minimum at the physical system ground 
state energy $E_0$ and electron number density $n_0({\mathbf r})$. 
The computational problem is to solve the KS equation 
\be
\lbrace -\half \nabla^2 + v_{\mathrm{KS}}([n];{\mathbf r})\rbrace \varphi_i({\mathbf r}) = \epsilon_i \varphi_i ({\mathbf r}) \; .
\label{ordinaryKS}
\ee
(in Hartree atomic units). The KS potential is 
\be
v_{\rm KS}=\delta (E_{\rm Ne} + E_{\rm H} + E_{\rm xc})/\delta n \equiv %
v_{\mathrm{ext}} + v_{\mathrm{H}} + v_{\mathrm{xc}} 
\label{KSpot}
\ee
where we have assumed, as appropriate for clamped nucleus
solids, that the external potential, $v_{\mathrm{ext}} = \delta E_{\rm Ne}/\delta n$, 
is from nuclear-electron attraction.  The electron-electron Coulomb
interaction energy customarily is partitioned as shown, namely the classical
Coulomb repulsion (Hartree energy), $E_{\mathrm H}$, and the residual 
exchange-correlation (XC) piece $E_{\mathrm {xc}}$.  Note that $E_{\mathrm {xc}}$ 
also contains the kinetic energy correlation 
contribution, namely the difference between the interacting and non-interacting
system kinetic energies ($T$ and $T_s$ respectively).  

The only term of this problem which is not known explicitly is $E_{\mathrm {xc}}$.
Great effort has gone into constructing approximations to it.  A convenient
classification, the Perdew-Schmidt Jacob's ladder \cite{PerdewSchmidt01}, 
proceeds by the number and type of ingredients, e.g. spatial derivatives,
non-interacting kinetic energy density, exact exchange, etc. For present
purposes the pertinent rungs of that ladder are the generalized gradient
approximation (GGA; dependent upon $n({\mathbf r})$ and $\nabla n({\mathbf r})$)
and meta-GGA functionals, which also depend upon the non-interacting KE 
density, 
\be
 t^{\mathrm{orb}}_{\rm s}({\mathbf r}) := \half \sum_{i=1}^{N_e}   %
|\nabla \varphi_i({\mathbf r})|^2   \; .
\label{KSKEden}
\ee
(For simplicity of exposition, we have assumed unit occupancy and
no degeneracy of KS orbitals.)  

Most often (but not universally) mGGA XC functionals use $t^{\mathrm{orb}}_{\rm s}$ 
in the combination \vspace{-4pt} 
\be
\alpha[n]:= (t_{\mathrm s}^{\mathrm {orb}}[n]-t_W[n])/t_{TF}[n] := t_\theta/t_{TF}
\label{alphadefn}
\vspace{-2pt}
\ee
as a way to detect chemically distinct spatial regions   
\cite{SunXiaoRuzsinszky12}.
The other ingredients in $\alpha[n]$ are the Thomas-Fermi \cite{Thomas,Fermi} 
and von Weizs\"acker \cite{Weizsacker} KE densities: \vspace{-4pt} 
\bea
t_{TF} &=& c_{TF} n^{5/3}({\mathbf r}) \;\;\;, %
\;\;\;  c_{TF} :=  \frac{3}{10}(3\pi^2)^{2/3} \label{ttf} \\
t_W &=& \frac{5}{3} t_{TF} s^2 \label{tw}  \; .
\eea
The dimensionless reduced density gradient used in GGAs and mGGAs is  \vspace{-4pt}
\be
s :=\frac{|\nabla n({\mathbf r})|}{2(3\pi^2)^{1/3}n^{4/3}({\mathbf r})} \; .
\label{sDefn}
\ee

Because $\alpha[n]$ is explicitly orbital dependent, the mGGA 
XC potential, 
$v_{\mathrm{xc}}=\delta E_{\mathrm{xc}}/\delta n$ is not calculable directly 
 but instead must be obtained as 
an optimized effective potential (OEP) 
\cite{StadeleMajewskiVoglGoerling1997,GraboKreibichGross1997,GraboKreibachKurthGross2000,HesselmannGoerling2008}.  The computational cost of 
OEP calculations is high enough that the procedure rarely is used 
in practice.   
Instead the so-called generalized Kohn-Sham (gKS) scheme is used.
In gKS, the variational procedure is done with respect to the
orbitals, not $n$. That delivers a set of  non-local
potentials $\delta E_{\mathrm{xc}}/\delta{\varphi_i}$ rather than the
local $v_{\mathrm{xc}}$.  For a mGGA, the KS and gKS schemes are inequivalent \cite{YangPengSunPerdewGKSBandGap2016}, a matter of both conceptual and practical 
consequences.  

Very recently we have shown \cite{DMRSBTPRA2017} that it is possible,
at least for molecules, 
to convert several successful mGGAs to Laplacian-level XC functionals,
mGGA-L, by a constraint-based deorbitalization strategy.  The scheme
is to evaluate $\alpha[n]$ with an \emph{orbital-independent}
approximation for $t_\theta$, i.e., for $t_{\mathrm s}^{\mathrm
  {orb}}[n]$. This is done with a KE density functional (KEDF) that is
parametrized to KS calculations on a small data set (18 atoms). The
parametrization is required to satisfy known constraints on the KE
density.  When tested against standard molecular datasets for a considerable
variety of properties, the deorbitalized (Laplacian-level)
versions of three well-known meta-GGAs, MVS \cite{MVS2015}, TPSS
\cite{TPSSa}, and SCAN \cite{SCAN}, gave as good or better results
than the originals.  Details are in Ref.\ \onlinecite{DMRSBTPRA2017}.

An obvious, crucial challenge is whether the \emph{identical} deorbitalization
of a mGGA can deliver equally satisfactory results on bulk solid validation
tests.  If that were to be true, then the deorbitalization strategy would 
be validated as truly successful in that 
it is general for ground states, not restricted to a particular state of 
aggregation.  Here we focus on the SCAN \cite{SCAN} functional because of
recent intense exploration of its broad efficacy on a considerable variety
of molecular and solid systems.  In short, we show that indeed SCAN-L, the 
deorbitalized SCAN, is essentially as accurate on a variety of solid
validation tests as the original SCAN. The deorbitalization strategy thus 
is validated as general, not specific to finite, self-bound systems.     

In order, the sections which follow give computational details
(Sect.\ \ref{ComputDetails}), numerical results (Sect.\ \ref{NumRes}), 
interpretive comparison of original and deorbitalized quantities such as $\alpha[n]$
(Sect.\ \ref{InterpRes}), computational performance (Sect.\ \ref{Comput}),
and brief conclusions (Sect.\ \ref{Concl}).

\section{\label{ComputDetails}Computational details}
	
The deorbitalized SCAN (SCAN-L) used in all calculations is precisely
the form and parametrization given in Ref.\ \onlinecite{DMRSBTPRA2017}.

All computations presented in this work were performed with a locally
modified version of the Vienna {\it ab initio} Simulation Package
(\textsc{vasp}).  Two separate implementations of the deorbitalized 
scheme were coded.  One used the mGGA trunk of \textsc{vasp}, modified
as necessary to handle the Laplacian-dependence.  That included use of
the KS rather than gKS solution.  The second version used the GGA trunk,
augmented to include the Laplacian in the one place it appears, $\alpha[n]$,
and its derivative appearance in $v_{\mathrm{KS}}$.  These coding differences
have pronounced computational performance consequences, as discussed
in Sect.\ \ref{Comput}.  

The PAW data sets utilized correspond to the PBE 5.4 package
and are summarized in Table \ref{PAWs}.  We note that the use of inconsistent 
PAW data sets (PBE with SCAN) follows precedent. To our
knowledge there is no alternative; no SCAN-based PAW data set is 
available for \textsc{vasp}. 
These PAW data sets contain
information about the core kinetic energy density needed by SCAN
\cite{SCAN}.  There are two exceptions, H and Li, for which the 
selected PAW data sets are all-electron but 
violate the requirement given in the VASP Wiki \cite{VASPwiki}. 
We found, nevertheless, that the equilibrium lattice constants for LiH, LiF, 
and LiCl 
from an equation of state fitted (see below) to calculations that used 
those PAW data sets are 
quite sensible. It is important to mention that in order to obtain the
same equilibrium lattice constants from the stress tensor values as from the
equation of state fitting 
the \emph{patch \#1} \cite{VASPpatch} needs to be applied to VASP.
	
The default energy cutoff (\textsc{vasp} variable {\tt ENCUT}) was 
overridden and set to 800 eV, except for calculations involving Li.  
In those, the cutoff was
increased to 1000 eV for LiCl and LiF, and to 1200 eV for Li. 
	
The precision parameter in \textsc{vasp} was set to ``accurate'' ({\tt PREC=A}) 
and the minimization algorithm used an ``all-band
simultaneous update of orbitals'' conjugate gradient method ({\tt
  ALGO=A}). Non-spherical contributions within the PAW spheres
were included self-consistently ({\tt LASPH=.TRUE.}).

For hexagonal close-packed structures we used the ideal 
$c/a$ ratio. For graphite and hexagonal boron nitride, we fixed the 
intralayer lattice constant to its experimental value and varied 
the interlayer lattice constant only.

Brillouin zone integrations were performed on ($17\times17\times17$)
$\Gamma$-centered symmetry reduced Monkhorst-Pack \cite{MonkhorstPack} 
$k$-meshes using the tetrahedron method with Bl\"ochl corrections 
\cite{Bloechl}. 	

The equilibrium lattice constants $a_0$ and bulk moduli $B_0$ at
$T=0K$ were determined by calculating the total energy per unit cell in
the range $V_0 \pm 10\%$ (where $V_0$ is the equilibrium unit cell
volume), followed by a twelve-point-fit to the stabilized jellium 
equation of state (SJEOS) \cite{SJEOS}. The SJEOS is
	\begin{equation}
	\label{eq:SJEOS}
	E(V) = \alpha \left( \frac{V_0}{V} \right) + \beta \left( \frac{V_0}{V} \right)^{2/3} + \gamma \left( \frac{V_0}{V} \right)^{1/3} + \omega \; .
	\end{equation}
A linear fit to Eq,\ (\ref{eq:SJEOS}) yields parameters
	$\alpha_s = \alpha V_0$, $\beta_s = \beta V_0^{2/3}$, 
	$\gamma_s = \gamma V_0^{1/3}$ and $\omega$, from which
	\begin{equation}
	 V_0 = \left( \frac{-\beta_s + \sqrt{\beta_s^2-3\alpha_s\gamma_s}}{\gamma_s} \right)^3 \; ,
	\end{equation}
	\begin{equation}
	B_0 = \frac{18\alpha + 10\beta + 4\gamma}{9V_0} \; .
	\end{equation}
	
To obtain cohesive energies, approximate isolated atom energies were
calculated from a $14\times15\times16$ \AA$^3$ cell.  The lowest
energy configuration was sought by allowing spin-polarization and
breaking spherical symmetry, but without spin-orbit coupling.  
	
\section{\label{NumRes}Numerical Results}
	
\begin{figure}[h]
	\includegraphics[width=0.8\linewidth]{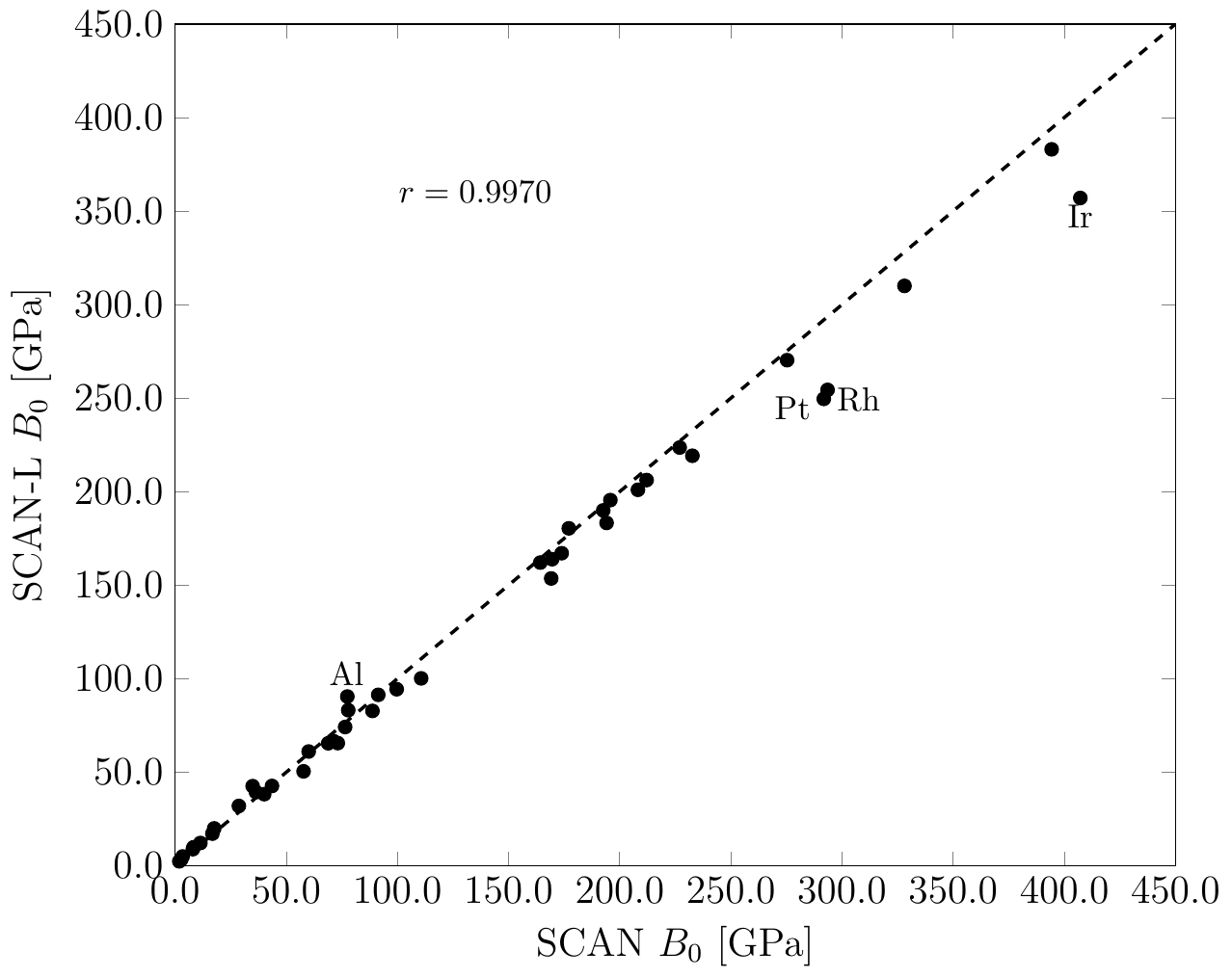}
	\includegraphics[width=0.8\linewidth]{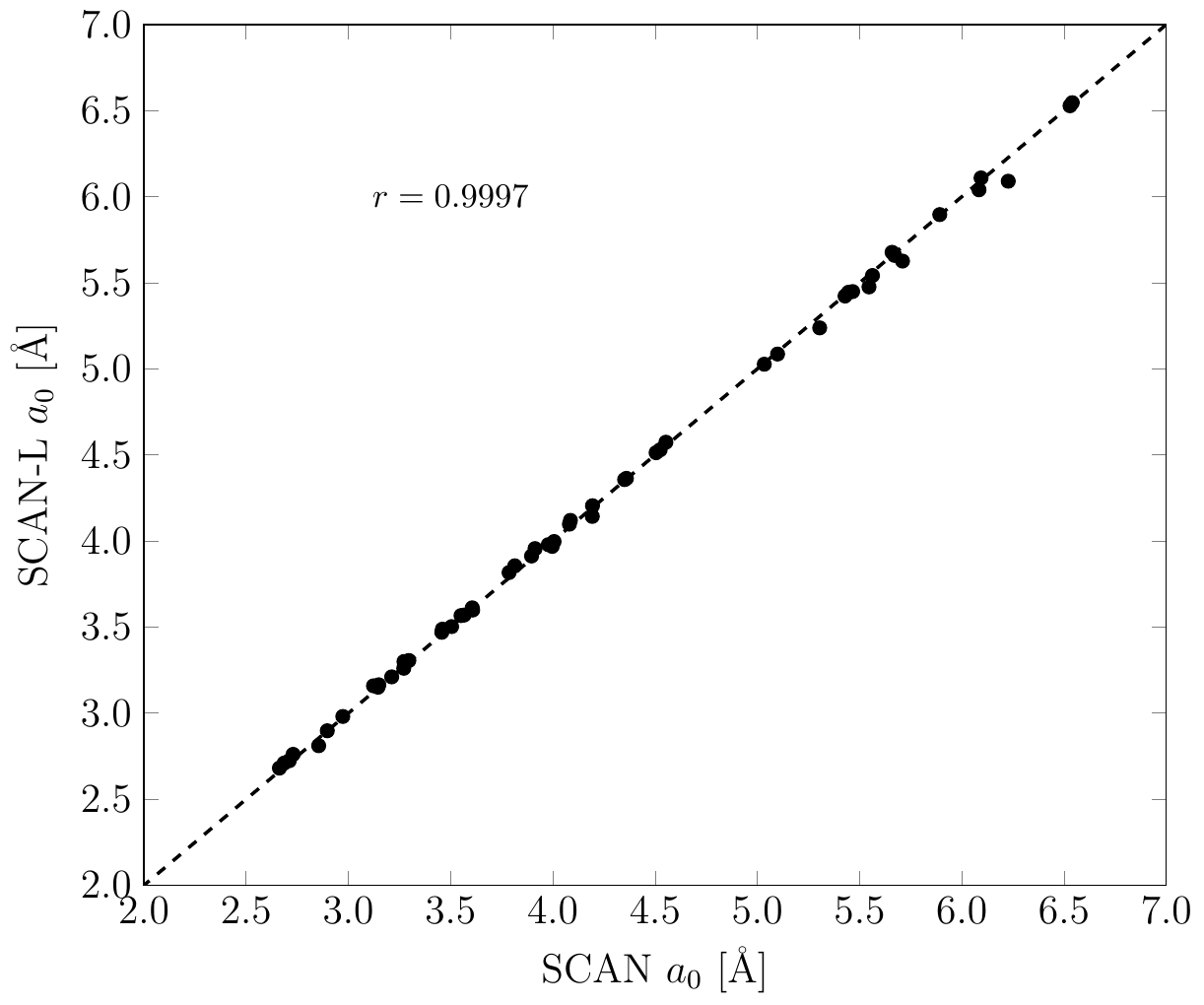}
	\includegraphics[width=0.8\linewidth]{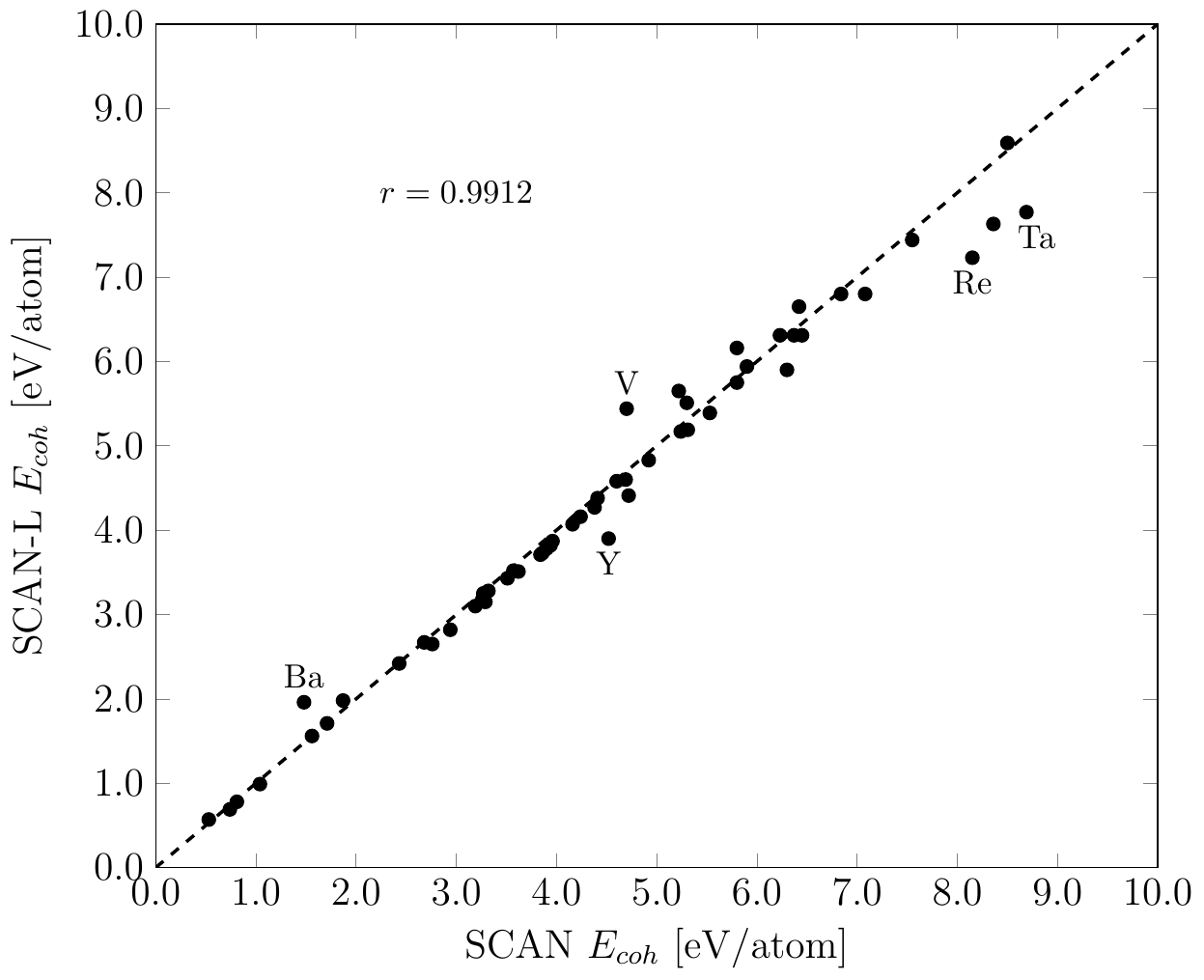}
	\caption{Comparison between SCAN and SCAN-L bulk moduli $B_0$ (top), static-crystal lattice constants $a_0$ (middle), and cohesive energies $E_{coh}$ (bottom). A dashed line with slope $1$ and Pearson correlation coefficient $r$ is shown in each of the three plots.\label{correlation}}
\end{figure}

Table \ref{a0-ecoh} compares static-crystal lattice
constants and cohesive energies of 55 solids and Table \ref{B0} 
compares bulk moduli of 44 solids computed with the orbital-dependent SCAN 
and its deorbitalized version SCAN-L. Experimental values shown in 
Table \ref{a0-ecoh} were taken from Ref.\ \onlinecite{SCAN+rVV10}. 
Those in Table \ref{B0} were taken from Ref.\ \onlinecite{TranStelzlBlaha}. 
The overall excellent agreement between the values obtained with SCAN and SCAN-L for all
three properties indicate that SCAN-L provides a \textit{faithful}
reproduction of the SCAN potential energy surfaces near equilibrium 
for these systems.
Figure \ref{correlation} depicts the correlation between SCAN and SCAN-L
results for each of the three properties. 

Outliers differing more than $\pm 10\%$ are indicated by their 
chemical symbol. It is readily apparent from Figure \ref{correlation}
that SCAN-L predicts Pt, Rh and Ir to be more compressible than does the
original SCAN functional. On the other hand, SCAN-L predicts Al, LiCl,
K, and Rb to be less compressible (these solids are not
highlighted in Figure \ref{correlation} due to cluttering).
At the resolution of that figure, there are 
no serious outliers for equilibrium lattice constant. In fact,
the differences between lattice constants predicted by SCAN and SCAN-L are 
$1\%$ or less for each one of the solids.  
There are 
a few outliers in the cohesive energies set. However, it is notable that 
there is no systematic under- or overbinding from SCAN-L with respect 
to SCAN values of $E_{coh}$. The ground-state configurations of all elements
were the same with SCAN and SCAN-L, however, we must note that one must start
from a previously converged PBE density to obtain the lowest lying state of
Hf with SCAN-L.

Table \ref{BG} displays KS band gaps for SCAN-L compared to those from SCAN.	
As expected, the SCAN-L band gap values always are less than or equal to 
those from the orbital-dependent SCAN. This systematic difference arises
because the exchange-correlation potential of SCAN-L is a local
multiplicative one, whereas the one from orbital-dependent SCAN is
non-multiplicative \cite{SCAN+BG}.  In other words, the difference in
KS band gaps is a direct
consequence of the difference between KS and generalized
gKS methods. If SCAN-L is a \textit{faithful} deorbitalization of
SCAN, then the SCAN-L exchange-correlation potential should be a 
good approximation to the SCAN OEP (which, so far as we know, has not
been generated for any solid),
Thus, the SCAN-L KS band gaps should agree reasonably well with the values
obtained in Ref.\ \onlinecite{SCAN+BG} for the Krieger-Li-Iafrate (KLI)
approximation to the OEP of SCAN. To facilitate such
comparison, Table \ref{BG} shows both the ``SCAN(BAND)'' and ``SCAN(KLI)''
band gaps reported in Ref.\ \onlinecite{SCAN+BG}.  ``SCAN(BAND)'' 
results are all-electron gKS values from the BAND code \cite{BANDcode}. 
The SCAN band gaps computed here with \textsc{vasp} and PAWs are smaller than those 
obtained with the BAND
code as reported in Ref. \onlinecite{SCAN+BG}.  Presumably that is 
a consequence of the PAWs and the difference in basis sets.  However, there is
no systematic deviation of the SCAN-L KS band gaps from the SCAN(KLI) ones. 
Most are close, with the two outliers, in a relative sense, being GaAs and InP: 
0.33 eV and 0.59 eV from SCAN-L versus 0.45 eV and 0.77 eV from SCAN(KLI), respectively.
It therefore seems that the SCAN-L potential is at least a 
reasonably good approximation to the SCAN OEP.
	
One of the features of SCAN that has been emphasized in the literature is 
its ability to capture intermediate-distance correlation effects in 
weakly bonded systems such as graphite and hexagonal boron nitride 
({\it h}-BN) \cite{SCAN+rVV10, SCANNature}. Table \ref{Eb} shows the
interlayer binding energy $E_b$ and interlayer lattice constant $c$ for these
two systems from SCAN and SCAN-L as well as reference values from Ref. \onlinecite{SCAN+rVV10}.
$E_b$ is small, thus particularly sensitive to formal and computational
differences.  Nevertheless SCAN-L reproduces the SCAN $E_b$ result for 
graphite and {\it h}-BN to less than 5.0\% discrepancy.  The SCAN-L $c$ values 
are somewhat closer to the reference ones but still agree to less than 2.5\% 
difference with 
the SCAN ones. Note, however, that our SCAN binding energies for
Graphite and {\it h}-BN are 10\% smaller than those reported in Reference [\onlinecite{SCAN+rVV10}].

\begin{figure}
	\includegraphics[width=0.8\linewidth]{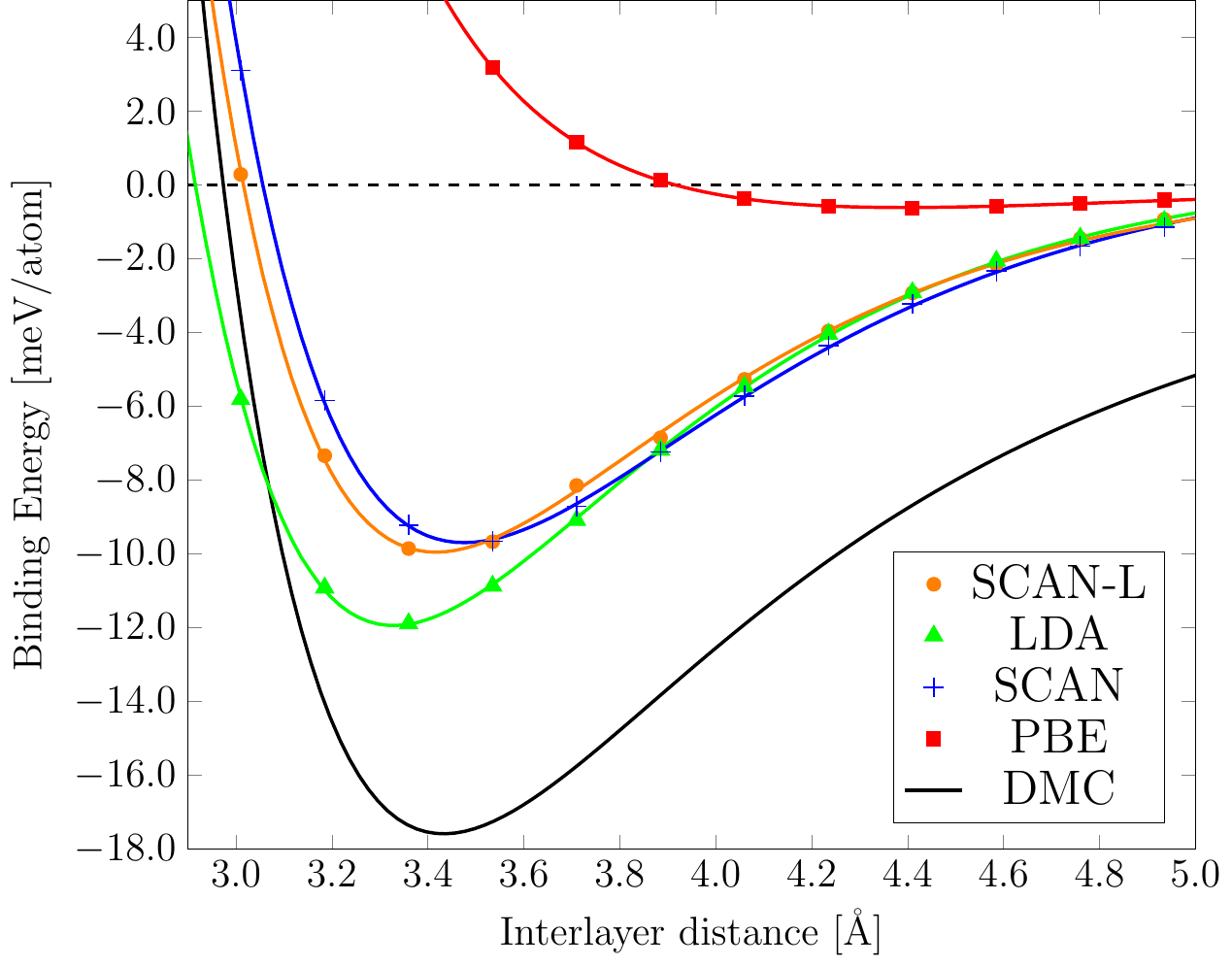}
	\caption{Graphene bilayer interplanar binding (meV/atom) as computed with LDA, PBE, SCAN, and SCAN-L XC functionals. A fit to diffusion quantum Monte Carlo (DMC) data \cite{DMC} is shown as reference. \label{fig:graphene}}
\end{figure}

It is also significant for use of van der Waals
corrections\cite{SCAN+rVV10,HermannTkatchenko} that SCAN-L reproduces
the SCAN binding curve $E_b(c)$ for a bilayer of graphene (BLG) quite
well. In Fig.\ \ref{fig:graphene} we show the binding curves for LDA,
PBE \cite{PBE}, SCAN and SCAN-L XC functionals compared to a diffusion
quantum Monte Carlo (DMC) reference \cite{DMC}.  The SCAN-L functional
is able to recover the binding ``lost'' by PBE and other GGA-type
functionals.  The interlayer separation predicted by SCAN (3.48 \AA)
and SCAN-L (3.41 \AA) are in good agreement to the DMC-fit prediction
(3.42 \AA).

	\begin{table}
	\caption{PAWs used in the present work. The energy cutoffs $E_{cut}$ (eV) shown are the default for each PAW and were overridden as discussed in the text.\label{PAWs}}
	\begin{tabular}{l l l l}\toprule
		Element & Name       & Valence      & $E_{cut}$     \\\hline
		H       & H\_h\_GW   & $1s$         & 700           \\			
		Li      & Li\_AE\_GW & $1s2s$       & 433           \\
		B       & B          & $2s2p$       & 319           \\
		C       & C\_GW\_new & $2s2p$       & 414           \\
		N       & N\_GW\_new & $2s2p$       & 421           \\
		O       & O\_GW\_new & $2s2p$       & 434           \\
		F       & F\_GW\_new & $2s2p$       & 487           \\
		Na      & Na\_pv     & $  2p3s$     & 373           \\
		Mg      & Mg\_sv\_GW & $2s2p3s$     & 430           \\
		Al      & Al\_GW     & $3s3p$       & 240           \\
		Si      & Si\_GW     & $3s3p$       & 547           \\
		P       & P\_GW      & $3s3p$       & 255           \\
		S       & S\_GW      & $3s3p$       & 259           \\
		Cl      & Cl\_GW     & $3s3p$       & 262           \\
		K       & K\_pv      & $3p4s$       & 249           \\
		Ca      & Ca\_pv     & $3p4s$       & 119           \\
		Sc      & Sc\_sv\_GW & $3s3p3d4s$   & 379           \\
		Ti      & Ti\_sv\_GW & $3s3p3d4s$   & 384           \\
		V       & V\_sv\_GW  & $3s3p3d4s$   & 382           \\
		Fe      & Fe\_sv\_GW & $3s3p3d4s$   & 388           \\
		Co      & Co\_sv\_GW & $3s3p3d4s$   & 387           \\
		Ni      & Ni\_sv\_GW & $3s3p3d4s$   & 390           \\
		Cu      & Cu\_GW     & $3d4s$       & 417           \\
		Ga      & Ga\_d\_GW  & $3d4s4p$     & 404           \\
		Ge      & Ge\_d\_GW  & $3d4s4p$     & 375           \\
		As      & As\_GW     & $4s4p$       & 208           \\
		Rb      & Rb\_sv     & $4s4p5s$     & 424           \\
		Sr      & Sr\_sv     & $4s4p5s$     & 229           \\
		Y       & Y\_sv\_GW  & $4s4p4d5s$   & 340           \\
		Zr      & Zr\_sv\_GW & $4s4p4d5s$   & 346           \\
		Nb      & Nb\_sv\_GW & $4s4p4d5s$   & 354           \\
		Mo      & Mo\_sv\_GW & $4s4p4d5s$   & 345           \\
		Tc      & Tc\_sv\_GW & $4s4p4d5s$   & 351           \\
		Ru      & Ru\_sv\_GW & $4s4p4d5s$   & 348           \\
		Rh      & Rh\_sv\_GW & $4s4p4d5s$   & 351           \\
		Pd      & Pd\_pv     & $4p4d5s$     & 250           \\
		Ag      & Ag\_GW     & $4d5s$       & 250           \\
		In      & In\_d\_GW  & $4d5s5p$     & 279           \\
		Sn      & Sn\_d\_GW  & $4d5s5p$     & 260           \\
		Sb      & Sb\_GW     & $5s5p$       & 172           \\
		Cs      & Cs\_sv\_GW & $5s5p6s$     & 198           \\
		Ba      & Ba\_sv\_GW & $5s5p6s$     & 237           \\
		Hf      & Hf\_sv\_GW & $5p6s6d$     & 283           \\
		Ta      & Ta\_sv\_GW & $5p6s6d$     & 286           \\
		W       & W\_sv\_GW  & $5p5d6s$     & 317           \\
		Re      & Re\_sv\_GW & $5p5d6s$     & 317           \\
		Os      & Os\_sv\_GW & $5p5d6s$     & 320           \\
		Ir      & Ir\_sv\_GW & $5p5d6s$     & 320           \\
		Pt      & Pt\_pv     & $5p5d6s$     & 295           \\
		Au      & Au\_GW     & $5d6s$       & 248           \\\toprule
	\end{tabular}
\end{table}

\begin{table}
	\caption{Strukturbericht symbols of the 57 solids used in the present work: A1 face-centered cubic, A2 body-centered cubic, A3 hexagonal close-packed, A9 hexagonal unbuckled graphite, B1 rock salt, B3 zinc blende, and
 B$_{k}$ hexagonal boron nitride. \label{Struktur}}
	\begin{tabular}{l l | l l | l l}\toprule
		Solid   &  Symbol & Solid   & Symbol  & Solid   & Symbol \\\hline
		C       &  A4     & NaF     &  B1     & Hf      &  A3 \\
		Si      &  A4     & NaCl    &  B1     & V       &  A2 \\
		Ge      &  A4     & MgO     &  B1     & Nb      &  A2 \\
		Sn      &  A4     & Li      &  A2     & Ta      &  A2 \\
		SiC     &  B3     & Na      &  A2     & Mo      &  A2 \\
		BN      &  B3     & K       &  A2     & W       &  A2 \\
		BP      &  B3     & Rb      &  A2     & Tc      &  A3 \\
		AlN     &  B3     & Cs      &  A2     & Re      &  A3 \\
		AlP     &  B3     & Ca      &  A1     & Ru      &  A3 \\
		AlAs    &  B3     & Ba      &  A2     & Os      &  A3 \\
		GaN     &  B3     & Sr      &  A1     & Rh      &  A1 \\
		GaP     &  B3     & Al      &  A1     & Ir      &  A1 \\
		GaAs    &  B3     & Fe      &  A2     & Pd      &  A1 \\
		InP     &  B3     & Co      &  A1     & Pt      &  A1 \\
		InAs    &  B3     & Ni      &  A1     & Cu      &  A1 \\
		InSb    &  B3     & Sc      &  A3     & Ag      &  A1 \\
		LiH     &  B1     & Y       &  A3     & Au      &  A1 \\
		LiF     &  B1     & Ti      &  A3     & C       &  A9 \\
		LiCl    &  B1     & Zr      &  A3     & BN      &  B$_k$
   \\\toprule
	\end{tabular}
\end{table}

\begin{table*}
	\caption{Static-lattice lattice constants, $a_0$ (\AA), and cohesive energies, $E_{coh}$ (eV/atom), of 55 solids. The experimental values,
from Ref.\ \onlinecite{SCAN+rVV10}, include zero-point effects. \label{a0-ecoh}}
	\begin{tabular}{l c c l c c l c c}\toprule
		\multirow{2}{*}{Solid} & \multicolumn{2}{c}{Experimental}& & \multicolumn{2}{c}{SCAN} & & \multicolumn{2}{c}{SCAN-L}  \\
		                       & $a_0$     & $E_{coh}$           & & $a_0$     & $E_{coh}$    & & $a_0$    & $E_{coh}$        \\\hline
		C                      &  3.553    &  7.55               & & 3.551     &  7.55        & & 3.567    & 7.44   \\ %
		Si                     &  5.421    &  4.68               & & 5.429     &  4.69        & & 5.423    & 4.60   \\ %
		Ge                     &  5.644    &  3.89               & & 5.668     &  3.94        & & 5.667    & 3.82   \\ %
		Sn                     &  6.477    &  3.16               & & 6.540     &  3.27        & & 6.546    & 3.25   \\ %
		SiC                    &  4.346    &  6.48               & & 4.351     &  6.45        & & 4.357    & 6.31   \\ %
		BN                     &  3.592    &  6.76               & & 3.606     &  6.84        & & 3.612    & 6.80   \\ %
		BP                     &  4.525    &  5.14               & & 4.525     &  5.31        & & 4.530    & 5.19   \\ %
		AlN                    &  4.368    &  5.85               & & 4.360     &  5.80        & & 4.364    & 5.75   \\ %
		AlP                    &  5.451    &  4.32               & & 5.466     &  4.24        & & 5.449    & 4.16   \\ %
		AlAs                   &  5.649    &  3.82               & & 5.671     &  3.84        & & 5.659    & 3.71   \\ %
		GaN                    &  4.520    &  4.55               & & 4.505     &  4.41        & & 4.513    & 4.38   \\ %
		GaP                    &  5.439    &  3.61               & & 5.446     &  3.62        & & 5.445    & 3.51   \\ %
		GaAs                   &  5.640    &  3.34               & & 5.659     &  3.29        & & 5.677    & 3.15   \\ %
		InP                    &  5.858    &  3.47               & & 5.892     &  3.19        & & 5.896    & 3.10   \\ %
		InAs                   &  6.047    &  3.08               & & 6.094     &  2.94        & & 6.109    & 2.82   \\ %
		InSb                   &  6.468    &  2.81               & & 6.529     &  2.68        & & 6.528    & 2.67   \\ %
		LiH                    &  3.979    &  2.49               & & 3.997     &  2.43        & & 3.969    & 2.42   \\ %
		LiF                    &  3.972    &  4.46               & & 3.978     &  4.38        & & 3.979    & 4.27   \\ %
		LiCl                   &  5.070    &  3.59               & & 5.099     &  3.51        & & 5.086    & 3.43   \\ %
		NaF                    &  4.582    &  3.97               & & 4.553     &  3.90        & & 4.574    & 3.78   \\ %
		NaCl                   &  5.569    &  3.34               & & 5.563     &  3.26        & & 5.542    & 3.18   \\ %
		MgO                    &  4.189    &  5.20               & & 4.194     &  5.24        & & 4.205    & 5.17   \\ %
		Li                     &  3.443    &  1.67               & & 3.457     &  1.56        & & 3.470    & 1.56   \\ %
		Na                     &  4.214    &  1.12               & & 4.193     &  1.04        & & 4.143    & 0.99   \\ %
		K                      &  5.212    &  0.94               & & 5.305     &  0.81        & & 5.238    & 0.78   \\ %
		Rb                     &  5.577    &  0.86               & & 5.710     &  0.74        & & 5.626    & 0.69   \\ %
		Cs                     &  6.039    &  0.81               & & 6.227     &  0.53        & & 6.090    & 0.57   \\ %
		Ca                     &  5.556    &  1.87               & & 5.546     &  1.87        & & 5.476    & 1.98   \\ %
		Ba                     &  5.002    &  1.91               & & 5.034     &  1.48        & & 5.027    & 1.96   \\ %
		Sr                     &  6.040    &  1.73               & & 6.084     &  1.71        & & 6.040    & 1.71   \\ %
		Al                     &  4.018    &  3.43               & & 4.006     &  3.57        & & 3.997    & 3.52   \\ %
		Fe                     &  2.853    &  4.30               & & 2.855     &  4.60        & & 2.811    & 4.58   \\ %
		Co                     &  3.524    &  4.42               & & 3.505     &  4.72        & & 3.503    & 4.41   \\ %
		Ni                     &  3.508    &  4.48               & & 3.460     &  5.30        & & 3.488    & 5.51   \\ %
		Sc                     &  3.270    &  3.93               & & 3.271     &  3.96        & & 3.261    & 3.87   \\ %
		Y                      &  3.594    &  4.39               & & 3.608     &  4.52        & & 3.599    & 3.90   \\ %
		Ti                     &  2.915    &  4.88               & & 2.897     &  4.92        & & 2.898    & 4.83   \\ %
		Zr                     &  3.198    &  6.27               & & 3.212     &  5.90        & & 3.211    & 5.94   \\ %
		Hf                     &  3.151    &  6.46               & & 3.123     &  6.30        & & 3.159    & 5.90   \\ %
		V                      &  3.021    &  5.35               & & 2.973     &  4.70        & & 2.981    & 5.44   \\ %
		Nb                     &  3.294    &  7.60               & & 3.296     &  6.37        & & 3.306    & 6.31   \\ %
		Ta                     &  3.299    &  8.13               & & 3.272     &  8.69        & & 3.300    & 7.77   \\ %
		Mo                     &  3.141    &  6.86               & & 3.145     &  5.80        & & 3.151    & 6.16   \\ %
		W                      &  3.160    &  8.94               & & 3.149     &  8.36        & & 3.165    & 7.63   \\ %
		Tc                     &  2.716    &  6.88               & & 2.711     &  6.42        & & 2.724    & 6.65   \\ %
		Re                     &  2.744    &  8.05               & & 2.730     &  8.15        & & 2.761    & 7.23   \\ %
		Ru                     &  2.669    &  6.77               & & 2.663     &  6.23        & & 2.681    & 6.31   \\ %
		Os                     &  2.699    &  8.20               & & 2.686     &  8.50        & & 2.710    & 8.59   \\ %
		Rh                     &  3.794    &  5.78               & & 3.786     &  5.22        & & 3.817    & 5.65   \\ %
		Ir                     &  3.831    &  6.99               & & 3.814     &  7.08        & & 3.856    & 6.80   \\ %
		Pd                     &  3.876    &  3.93               & & 3.896     &  4.16        & & 3.913    & 4.07   \\ %
		Pt                     &  3.913    &  5.87               & & 3.913     &  5.53        & & 3.956    & 5.39   \\ %
		Cu                     &  3.595    &  3.51               & & 3.566     &  3.86        & & 3.570    & 3.73   \\ %
		Ag                     &  4.062    &  2.96               & & 4.081     &  2.76        & & 4.098    & 2.65   \\ %
		Au                     &  4.062    &  3.83               & & 4.086     &  3.32        & & 4.120    & 3.28   \\\hline %
		ME                     &           &                     & & 0.011     & -0.10        & & 0.009    &-0.17   \\
		MAE                    &           &                     & & 0.025     &  0.24        & & 0.024    & 0.26   \\
		MARE (\%)              &           &                     & & 0.54      &  5.93        & & 0.55     & 6.42   \\\toprule
	\end{tabular}
\end{table*}

	\begin{table}
	\caption{Bulk modulus, $B_0$ (GPa) of the 44 cubic solids. The experimental values, from Ref.\ \onlinecite{TranStelzlBlaha}, were obtained by subtracting the zero-point phonon effect from the experimental zero-temperature values. \label{B0}}
	\begin{tabular}{l c c c }\toprule
		Solid  &  Expt.   & SCAN  & SCAN-L \\\hline
		C      &  454.7   & 459.9 & 442.6  \\
		Si     &  101.3   &  99.7 &  94.4  \\
		Ge     &   79.4   &  71.2 &  66.7  \\
		Sn     &   42.8   &  40.1 &  38.3  \\
		SiC    &  229.1   & 227.0 & 223.6  \\
		BN     &  410.2   & 394.3 & 383.0  \\
		BP     &  168.0   & 173.9 & 167.1  \\
		AlN    &  206.0   & 212.1 & 206.2  \\
		AlP    &   87.4   &  91.4 &  91.4  \\
		AlAs   &   75.0   &  76.5 &  74.2  \\
		GaN    &  213.7   & 194.1 & 183.3  \\
		GaP    &   89.6   &  88.8 &  82.8  \\
		GaAs   &   76.7   &  73.2 &  65.6  \\
		InP    &   72.0   &  68.9 &  65.5  \\
		InAs   &   58.6   &  57.8 &  50.5  \\
		InSb   &   46.1   &  43.6 &  42.7  \\
		LiH    &   40.1   &  36.4 &  39.4  \\
		LiF    &   76.3   &  77.9 &  83.2  \\
		LiCl   &   38.7   &  34.9 &  42.6  \\
		NaF    &   53.1   &  60.1 &  61.1  \\
		NaCl   &   27.6   &  28.7 &  32.0  \\
		MgO    &  169.8   & 169.6 & 163.9  \\
		Li     &   13.1   &  16.8 &  17.2  \\
		Na     &    7.9   &   8.0 &   8.9  \\
		K      &    3.8   &   3.4 &   5.0  \\
		Rb     &    3.6   &   2.7 &   3.3  \\
		Cs     &    2.3   &   1.9 &   2.4  \\
		Ca     &   15.9   &  17.6 &  20.0  \\
		Ba     &   10.6   &   8.3 &   9.9  \\
		Sr     &   12.0   &  11.4 &  12.2  \\										
		Al     &   77.1   &  77.5 &  90.5  \\
		Ni     &  192.5   & 232.7 & 219.2  \\
		V      &  165.8   & 195.8 & 195.5  \\		
		Nb     &  173.2   & 177.1 & 180.4  \\
		Ta     &  202.7   & 208.2 & 201.0  \\
		Mo     &  276.2   & 275.3 & 270.3  \\
		W      &  327.5   & 328.1 & 310.0  \\
		Rh     &  277.1   & 293.5 & 254.4  \\
		Ir     &  362.2   & 407.2 & 357.0  \\
		Pd     &  187.2   & 192.6 & 190.0  \\
		Pt     &  285.5   & 291.8 & 249.6  \\		
		Cu     &  144.3   & 164.3 & 162.1  \\
		Ag     &  105.7   & 110.7 & 100.2  \\
		Au     &  182.0   & 169.2 & 153.6  \\\hline
		ME     &          & 3.0   & -3.0    \\
		MAE    &          & 6.9   &  9.2    \\
		MARE (\%) &       & 7.1   &  9.4   \\\toprule
	\end{tabular}
\end{table}

	\begin{table*}
	\caption{Band gap (eV) of 21 insulators and semiconductors. Experimental lattice parameters were used with all functionals. Experimental band gaps and lattice constants were taken from Ref.\ \onlinecite{TranBlahaBG}.\label{BG}}
	\begin{tabular}{c c c c c c}\toprule
		Solid  &  Expt.   & SCAN   & SCAN-L & SCAN(BAND) & SCAN(KLI) \\\hline
		C      &  5.50    & 4.54   & 4.22   & 4.58       & 4.26 \\
		Si     &  1.17    & 0.82   & 0.80   & 0.97       & 0.78 \\
		Ge     &  0.74    & 0.14   & 0.00   &            &      \\
		SiC    &  2.42    & 1.72   & 1.55   &            &      \\
		BN     &  6.36    & 4.98   & 4.66   & 5.04       & 4.73 \\
		BP     &  2.10    & 1.54   & 1.41   & 1.74       & 1.52 \\
		AlN    &  4.90    & 3.97   & 3.50   &            &      \\
		AlP    &  2.50    & 1.92   & 1.81   &            &      \\
		AlAs   &  2.23    & 1.74   & 1.59   &            &      \\
		GaN    &  3.28    & 1.96   & 1.49   &            &      \\
		GaP    &  2.35    & 1.83   & 1.72   & 1.94       & 1.72 \\
		GaAs   &  1.52    & 0.77   & 0.33   & 0.80       & 0.45 \\
		InP    &  1.42    & 1.02   & 0.59   & 1.06       & 0.77 \\
		InAs   &  0.42    & 0.00   & 0.00   &            &      \\
		InSb   &  0.24    & 0.00   & 0.00   &            &      \\
		LiH    &  4.94    & 3.66   & 3.69   &            &      \\
		LiF    &  14.20   & 10.10  & 9.16   & 9.97       & 9.11 \\
		LiCl   &  9.40    & 7.33   & 6.80   &            &      \\
		NaF    &  11.50   & 7.14   & 6.45   &            &      \\
		NaCl   &  8.50    & 5.99   & 5.59   & 5.86       & 5.25 \\
		MgO    &  7.83    & 5.79   & 4.92   & 5.62       & 4.80 \\\hline
		ME     &          & -1.26  & -1.58  &            &      \\
		MAE    &          &  1.26  &  1.58  &            &      \\\toprule
	\end{tabular}
\end{table*}

\begin{table}[h]
	\caption{Inter-layer binding energy $E_b$ in meV/\AA$^2$ and inter-layer lattice constant $c$ in \AA. The reference  $E_b$ values  are from RPA calculations and from experiments for $c$.\label{Eb}}
	\begin{tabular}{c c c c c c c}\toprule
		\multirow{2}{*}{Solid} & \multicolumn{2}{c}{Reference} & \multicolumn{2}{c}{SCAN} & \multicolumn{2}{c}{SCAN-L} \\
		                       &  $E_b$  &  $c$  & $E_b$  &  $c$  & $E_b$  &  $c$  \\\hline
		Graphite               &  18.32  &  6.70 &  7.23  &  6.97 &  7.37  & 6.81  \\
		$h$-BN                 &  14.49  &  6.54 &  7.66  &  6.85 &  7.70  & 6.72  \\\toprule
	\end{tabular}
\end{table}

\begin{table}[h]
	\caption{Comparative timings for PBE, SCAN, and SCAN-L 
calculations in the original and modified  mGGA and GGA trunks of
\textsc{vasp}.  All times in seconds. See text for trunk labels. \label{timings}}
	\begin{tabular}{l c c c}\toprule
	\multirow{2}{*}{XC} & \multirow{2}{*}{Trunk} & Original & Modified \\
	 &  &  Code &  Code \\ \hline 
       PBE & GGA=PE       &  12.38  &  12.85  \\
       PBE & METAGGA=PBE   &  36.75  &  37.57  \\
       SCAN & METAGGA=SCAN  &  61.28 & --\\
       SCAN-L & GGA=SL        &  --    & 19.32 \\
       SCAN-L & METAGGA=SCANL &  --    & 50.72 \\ \toprule 
	\end{tabular}
\end{table}

\section{\label{InterpRes} Interpretive Results}

\begin{figure}
	\includegraphics[width=0.9\linewidth]{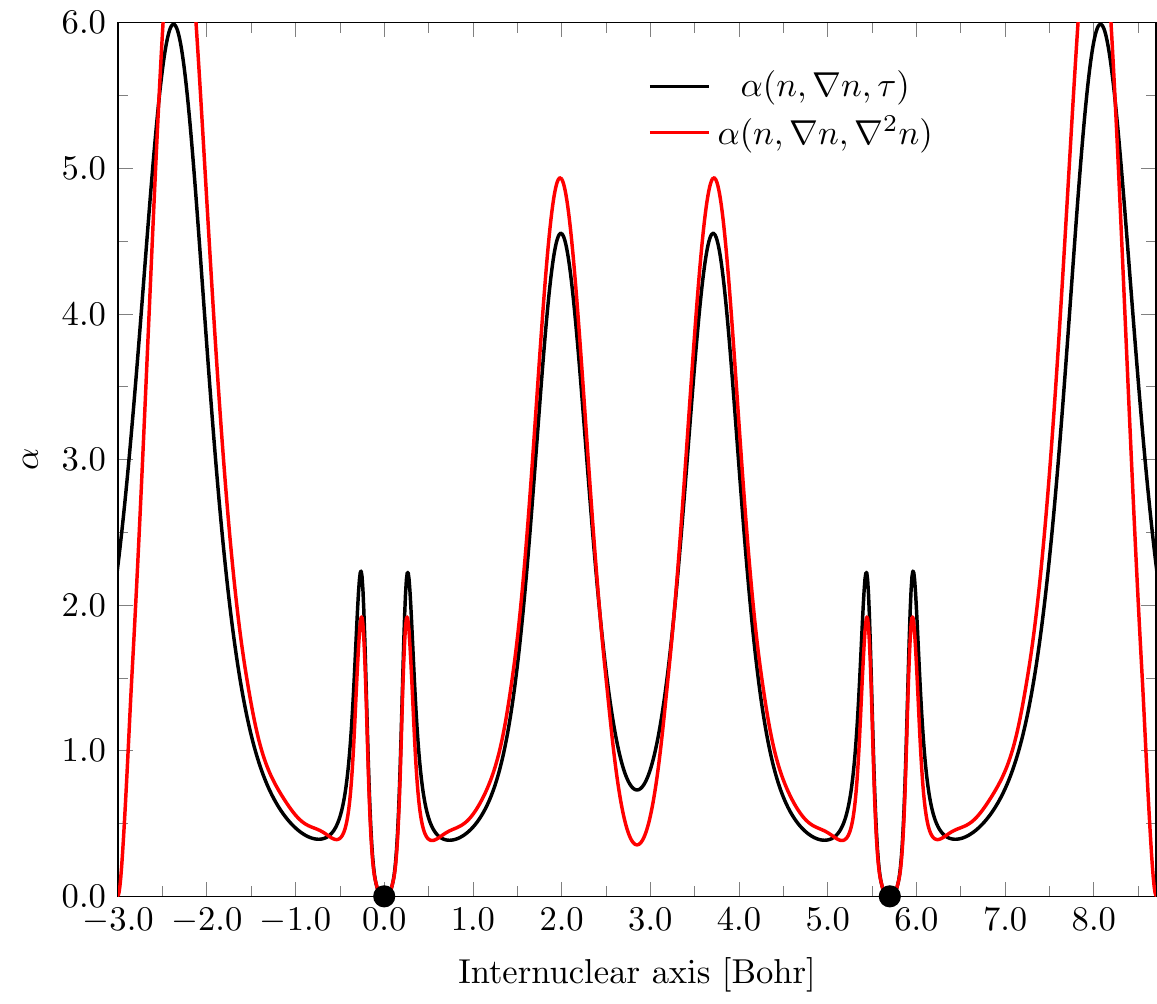}
	\includegraphics[width=0.9\linewidth]{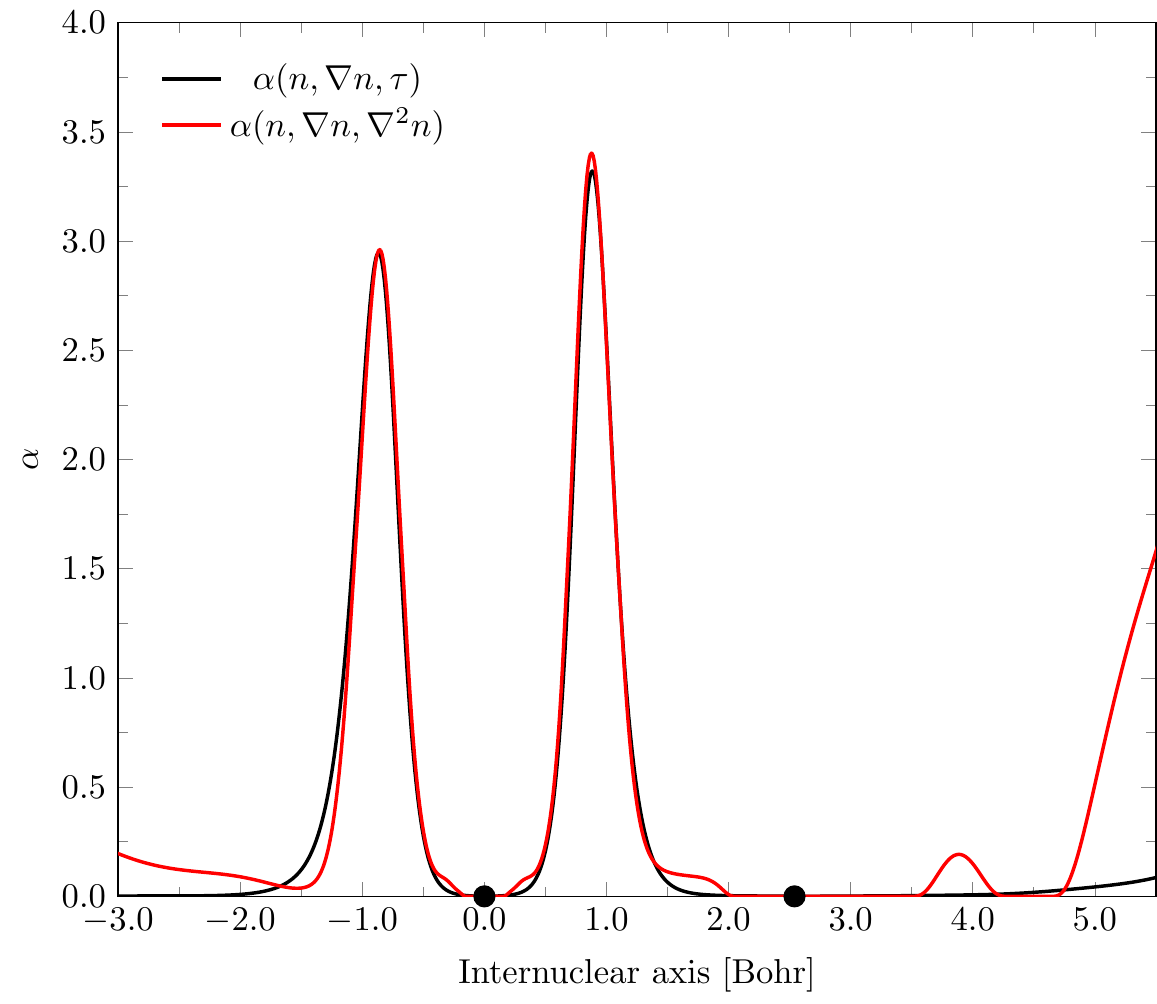}
	\includegraphics[width=0.9\linewidth]{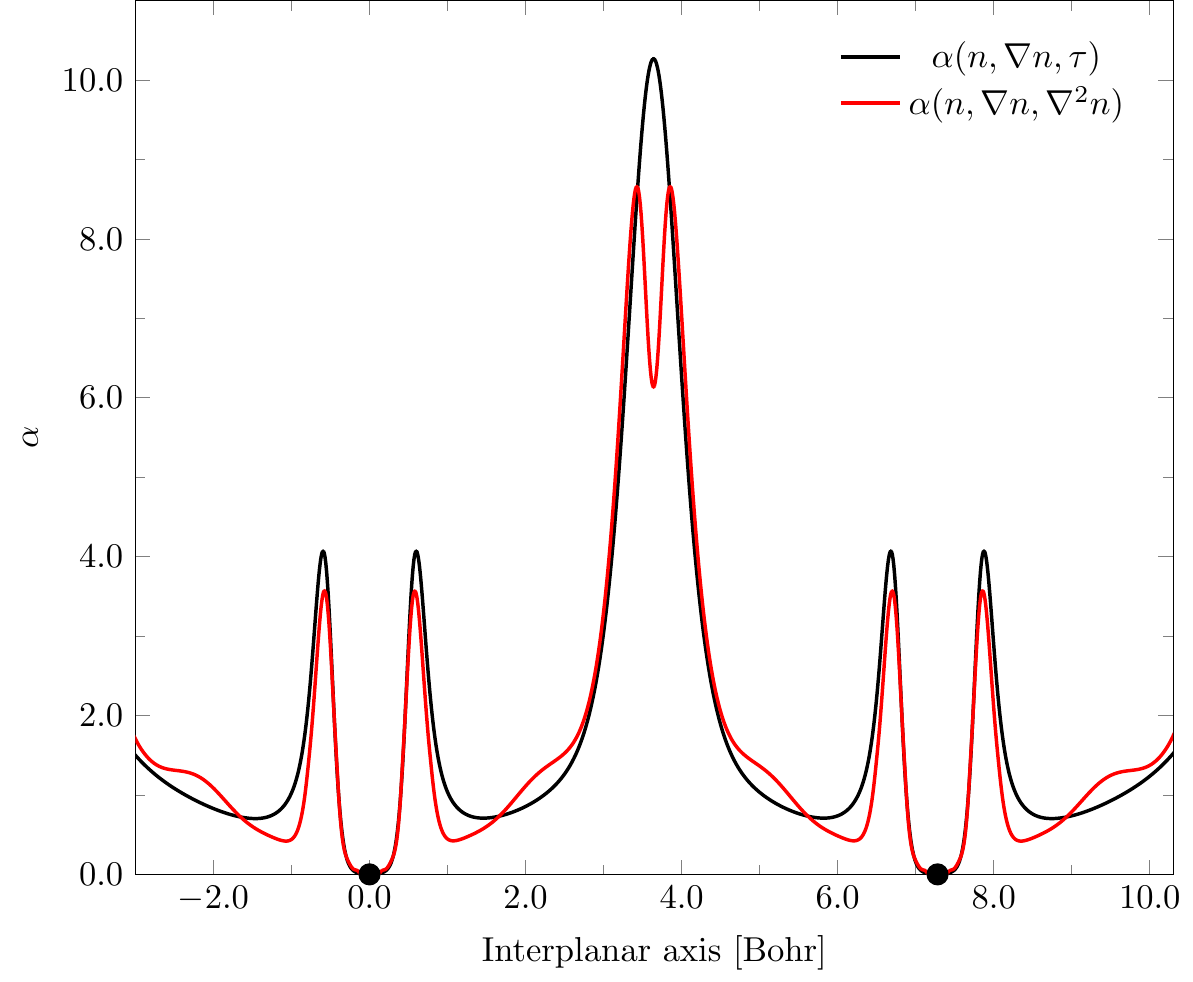}
	\caption{Orbital-dependent $\alpha$ and its deorbitalized approximation for three representative bonding situations. Na$_2$ (top) exemplifies systems with $\alpha \approx 1$; BeH (middle) exemplifies systems with $\alpha \approx 0$, and the benzene dimer exemplifies those with $\alpha \gg 1$. \label{alpha}}
\end{figure}

How the  \emph{faithful} deorbitalization is achieved in  SCAN-L
can be understood by how well $\alpha$ is approximated
by the kinetic energy density functional utilized. Figure \ref{alpha} 
shows the comparison
between the orbital-dependent $\alpha$ Eq.\ (\ref{alphadefn}) and the
approximation that results from use of the modified PC07 kinetic energy density
functional in deorbitalizing SCAN \cite{DMRSBTPRA2017}.  Details of the
parametrization of PC07 are in that reference.  
The three systems selected as examples in that Figure were chosen because they 
span the bonding situations among which $\alpha$ is supposed to 
discriminate.  The BeH radical has $\alpha \approx 0$ as is true of 
most covalently bounded systems. The sodium dimer has
$\alpha \approx 1$ as in most metallic systems. The stacked benzene dimer
is representative of weakly bound systems for which $\alpha \gg 1$ is typical.
The deorbitalized $\alpha$ follows its orbital-dependent
counterpart closely both within and outside the bonding regions. Important differences
can be noted for the benzene dimer at the mid-point 
of the interplanar axis.  However, even though the deorbitalized 
$\alpha$ is almost 50 \% of the exact one, the difference between SCAN 
and SCAN-L enhancement factors is less than 5 \%.
Larger differences between the exact and approximate $\alpha$s might be 
observed in the tails of the density.  Those are almost nonexistent in 
condensed systems near equilibrium and they prove to be inconsequential
for molecules (which is why such points are screened out in most
molecular computational packages).  In short, where it counts in 
both solids and molecules, the PC07 function reproduces
the behavior of the original SCAN $\alpha$.

\section{\label{Comput}Computational Performance}

To obtain a quantitative picture of the performance of GGA and mGGA
calculations in \textsc{vasp}, we prepared a fully sequential (serial)
version compiled in profiling mode and linked to the Intel Math Kernel
and Fast Fourier Transform libraries.  Two variants were compiled,
original and with SCAN-L included.  Within the original variant, three
calculations were done: PBE using the GGA trunk ({\tt GGA=PBE}), 
PBE using the mGGA trunk ({\tt METAGGA=PBE}) and SCAN ({\tt METAGGA=SCAN}). 
Correspondingly in the
SCAN-L coded variant, the four were PBE using the extended GGA
trunk ({\tt GGA=PE}), PBE using the modified mGGA trunk ({\tt METAGGA=PBE}), 
SCAN-L using the extended GGA trunk ({\tt GGA=SL}) and 
SCAN-L using the modified mGGA trunk ({\tt METAGGA=SCANL}).

The test system was diamond carbon at a lattice constant, $a_0$ = 3.560 \AA,
near the SCAN-L equilibrium value. Calculations used a 600 eV
planewave cutoff, the all-bands conjugate gradient minimization
({\tt ALGO=A}), aspherical corrections ({\tt LASPH=.TRUE.}), 
doubling of the Fourier grid ({\tt ADDGRID=.TRUE.})
and the tetrahedron method with Bl\"ochl corrections ({\tt ISMEAR=-5}). These
are the same settings as were used for the validation studies, except for
the energy cutoff. All calculations converged in 12 scf iterations.

Table \ref{timings} gives the results.  Clearly the SCAN-L computational 
speed in the extended GGA trunk implementation is substantially superior to 
that for original SCAN. When using the mGGA trunk, SCAN-L performance degrades
but the computation is still 20\% faster than the original SCAN one.

Analysis of the detailed profiling shows that 
when a metaGGA functional is requested, \textsc{vasp} first 
computes results for PBE XC, but then
overwrites those results with the corresponding ones from the 
requested  mGGA XC functional. It is not clear why that is done.  For
mGGAs, additional time is used in computing $\nabla^2 n$, 
especially on the radial grid within the PAW
spheres, even if that Laplacian data actually is un-needed
in the requested mGGA XC functional. (Of course, it is
used in SCAN-L.)  That is done
in anticipation of computing the modified Becke-Johnson potential (also
called Tran-Blaha 09)\cite{BJ,TB09} 
if requested.   Furthermore, the mGGA trunk always assumes spin-polarized
densities, resulting in additional time used for spin-unpolarized systems.
These three sources of wasted time make the difference between
the {\tt GGA=PE} and {\tt METAGGA=PBE} timings.

In the original \textsc{vasp} version, the time 
difference between
{\tt METAGGA=PBE} and {\tt METAGGA=SCAN} arises from
the non-locality of the Hamiltonian of a conventional mGGA (as in SCAN).
Implementing SCAN-L as an extension of the GGA trunk saves time because the
associated $v_{\mathrm{KS}}$ is
local. That implementation also avoids wasting time calculating 
un-needed PBE results and avoids treating spin unpolarized systems
as spin polarized ones.

\section{\label{Concl}Conclusion}

We have shown that the SCAN-L functional, a simple orbital-independent
form of the sophisticated and much-advertised SCAN functional, can
capture all the pertinent details of orbital-dependent functionals
both in molecules and solids. We believe SCAN-L to be the first example of an
orbital-independent functional that provides uniformly rather good
performance in these two seemingly irreconcilable  domains of
aggregation.  As such, SCAN-L opens the way for meta-GGA XC 
accuracy and reliability in orbital-free DFT simulations, a 
possibility that has not existed heretofore.  It also opens the
way for much faster ab initio molecular dynamics simulations than
are possible with SCAN.  

Differences between SCAN-L and SCAN KS band gaps arise as a 
well-understood consequence of the difference between KS and gKS 
solutions.  The KS band gaps also provide some evidence that SCAN-L provides
a reasonable approximation to the OEP for SCAN.  Direct comparison with
the exact OEP (rather than the KLI approximation) would be welcome. 
The performance of SCAN-L in combination with van der Waals 
correction schemes also remains to be investigated.

\begin{acknowledgments}
This work was supported  by U.S.\ National Science Foundation 
grant DMR-1515307, and, in the last phases, by U.S.\ Dept.\ of
Energy grant DE-SC 0002139.  
\end{acknowledgments}

\end{document}